\documentclass[conference]{IEEEtran}
\IEEEoverridecommandlockouts
\usepackage{cite}
\usepackage{amsmath,amssymb,amsfonts,upgreek}
\usepackage{algorithmic}
\usepackage{algorithm}
\usepackage{graphicx}
\usepackage{textcomp}
\usepackage{xcolor}
\usepackage{stfloats}

\usepackage[left=0.525in,right=0.54in,top=0.76in, bottom=0.935in]{geometry}

\usepackage{hhline}
\usepackage{makecell}
\usepackage{booktabs}
\usepackage[flushleft]{threeparttable}

\def\BibTeX{{\rm B\kern-.05em{\sc i\kern-.025em b}\kern-.08em
    T\kern-.1667em\lower.7ex\hbox{E}\kern-.125emX}}
\begin{document}

\title{On the Parameter Selection of Phase-transmittance Radial Basis Function Neural Networks \\ for Communication Systems}


\author{\IEEEauthorblockN{Jonathan A. Soares, Kayol S. Mayer, and Dalton S. Arantes}
\IEEEauthorblockA{\textit{Department of Communications, School of Electrical and Computer Engineering} \\
\textit{Universidade Estadual de Campinas -- UNICAMP}, Campinas, SP, Brazil\\
j229966@dac.unicamp.br, kayol@unicamp.br, dalton@unicamp.br}
}

\maketitle

\begin{abstract}
In the ever-evolving field of digital communication systems, complex-valued neural networks (CVNNs) have become a cornerstone, delivering exceptional performance in tasks like equalization, channel estimation, beamforming, and decoding. Among the myriad of CVNN architectures, the phase-transmittance radial basis function neural network (PT-RBF) stands out, especially when operating in noisy environments such as 5G MIMO systems. Despite its capabilities, achieving convergence in multi-layered, multi-input, and multi-output PT-RBFs remains a daunting challenge. Addressing this gap, this paper presents a novel Deep PT-RBF parameter initialization technique. Through rigorous simulations conforming to 3GPP TS 38 standards, our method not only outperforms conventional initialization strategies like random, $K$-means, and constellation-based methods but is also the only approach to achieve successful convergence in deep PT-RBF architectures. These findings pave the way to more robust and efficient neural network deployments in complex digital communication systems.
\end{abstract}

\section{Introduction}

Recently, in communication systems, complex-valued neural networks~(CVNNs) have been studied in several applications, such as equalization, channel estimation, beamforming, and decoding~\cite{Mayer2019a,Ding2020,Zhang2021b,Li2022,Mayer2020a,Kamiyama2021,Freire2020,Soares2021,Xu2022, Chu2022,Mayer2022,Yang2022b,Xiao2023}. This growing interest is related to enhanced functionality, improved performance, and reduced training time when compared with real-valued neural networks~(RVNNs)~\cite{Hirose2012b, Barrachina2021, Cruz2022, Zhang2022}.

The effectiveness of neural networks is critically dependent on several factors, such as initialization, regularization, and optimization~\cite{Humbird2019}. Although regularization and optimization techniques are vital to speed up the training process and reduce steady-state error~\cite{Hu2021}, depending on the initial parameter selection, the neural network can get stuck at local minima, achieving suboptimal solutions~\cite{Narkhede2022}. For radial basis function~(RBF)-based neural networks, this problem is even worse since, for each layer, there are four parameters~(synaptic weight, bias, center vectors, and center variances) in contrast to two parameters~(synaptic weights and bias) of usual multilayer perceptron neural networks.


In this context, with a focus on the phase-transmittance radial basis function~(PT-RBF) neural network~\cite{Mayer2022}, we propose a novel parameter selection scheme. This scheme aims to initialize synaptic weights, biases, center vectors, and center variances in the complex domain. Notably, existing literature offers limited guidance on initialization techniques for multilayer RBF-based CVNNs. Despite this gap, our study compares the proposed approach against well-known methods such as random initialization~\cite{Wallace2005}, K-means clustering~\cite{Turnbull2005}, and constellation-based initialization~\cite{Loss2007}. To the best of our knowledge, this is the first work that handles this initialization challenge for multi-layered PT-RBFs.


\section{Complex-valued PT-RBF Neural Networks}
\label{sec:cvnns}

Following the notation used by \cite{Mayer2022}, the deep PT-RBF is defined with $L$ hidden layers~(excluding the input layer), where the superscript $l \in [0,\,1,\,\cdots,\,L]$ denotes the layer index and $l=0$ is the input layer. The $l$-th layer~(excluding the input layer $l=0$) is composed by $I^{\{l\}}$ neurons, $O^{\{l\}}$ outputs, and has a matrix of synaptic weights $\mathbf{W}^{\{l\}}\in\mathbb{C}^{O^{\{l\}}\times I^{\{l\}}}$, a bias vector $\mathbf{b}^{\{l\}}\in\mathbb{C}^{O^{\{l\}}}$, a matrix of center vectors $\boldsymbol{\Gamma}^{\{l\}}\in\mathbb{C}^{I^{\{l\}}\times O^{\{l-1\}}}$, and a variance vector $\boldsymbol{\upsigma}^{\{l\}}\in\mathbb{C}^{I^{\{l\}}}$. Notice that $\mathbf{\bar{x}}\in \mathbb{C}^P$ is the deep PT-RBF normalized input vector~($P$ inputs) and $\mathbf{y}^{\{L\}}\in \mathbb{C}^R$ is the deep PT-RBF output vector~($R$ outputs). The $l$-th hidden layer output vector $\mathbf{y}^{\{l\}} \in \mathbb{C}^{O^{\{l\}}}$ is given by
\begin{equation}
\label{eq:deepPTRBFNN_output}
    \mathbf{y}^{\{l\}}=\mathbf{W}^{\{l\}}\boldsymbol{\upphi}^{\{l\}}+\mathbf{b}^{\{l\}}, 
\end{equation}
where $\boldsymbol{\upphi}^{\{l\}}\in \mathbb{C}^{I^{\{l\}}}$ is the vector of Gaussian kernels.

The $m$-th Gaussian kernel of the $l$-th hidden layer is formulated as
\begin{equation}
\label{eq:kernel}
\phi_m^{\{l\}}=\exp\left[-\Re\left(v_m^{\{l\}}\right)\right]+\jmath\exp\left[-\Im\left(v_m^{\{l\}}\right)\right],      
\end{equation}
in which $v_m^{\{l\}}$ is the $m$-th Gaussian kernel input of the $l$-th hidden layer, described as
\begin{multline}
\label{eq:kernel_argument}
v_m^{\{l\}}=\frac{\left \Vert\Re\left(\mathbf{y}^{\{l-1\}}\right)-\Re\left(\boldsymbol{\upgamma}_m^{\{l\}}\right)\right \Vert_2^2}{\Re\left(\sigma_m^{\{l\}}\right)} \\
+\jmath\frac{\left \Vert\Im\left(\mathbf{y}^{\{l-1\}}\right)-\Im\left(\boldsymbol{\upgamma}_m^{\{l\}}\right)\right \Vert_2^2}{\Im\left(\sigma_m^{\{l\}}\right)},
\end{multline}
where $\mathbf{y}^{\{l-1\}}\in \mathbb{C}^{O^{\{l-1\}}}$ is the output vector of the $(l-1)$-th hidden layer~(except for the first hidden layer that $\mathbf{y}^{\{0\}}=\mathbf{\bar{x}}$), $\boldsymbol{\upgamma}_m^{\{l\}}\in \mathbb{C}^{O^{\{l-1\}}}$ is the $m$-th vector of Gaussian centers of the $l$-th hidden layer, $\sigma_m^{\{l\}}\in\mathbb{C}$ is the respective $m$-th variance, and $\Re(\cdot)$ and $\Im(\cdot)$ return the real and imaginary components, respectively.

\section{Initialization of Complex-valued Radial Basis Function Neural Networks}

\subsection{Random Initialization}
Based on real-valued initialization~\cite{Wallace2005}, one of the simplest and easiest ways to initialize the parameters of a complex-valued RBF-based neural network is setting~$\boldsymbol{\Gamma}^{\{l\}}$ and $\mathbf{W}^{\{l\}}$ randomly, as
\begin{equation}
    \boldsymbol{\Gamma}^{\{l\}} \sim \mathcal{C}\mathcal{G}\left(0,\sigma^2_{\boldsymbol{\Gamma}^{\{l\}}} \right),
\end{equation}
\begin{equation}
    \mathbf{W}^{\{l\}} \sim \mathcal{C}\mathcal{G}\left(0,1 \right),
\end{equation}
in which $\mathcal{C}\mathcal{G}(\cdot)$ is a generic complex-valued distribution function, and $\sigma^2_{\boldsymbol{\Gamma}^{\{l\}}}$ is the desired variance of $\boldsymbol{\Gamma}^{\{l\}}$.

On the other hand, the bias and center variances are initialized as constant values
\begin{equation}
    \mathbf{b}^{\{l\}}=\mathbf{0}+\jmath\mathbf{0},
\end{equation}
\begin{equation}
    \boldsymbol{\upsigma}^{\{l\}}=\frac{\sigma^2_{\boldsymbol{\Gamma}^{\{l\}}}}{2}\left(\mathbf{1}+\jmath\mathbf{1}\right),
\end{equation}
where $\mathbf{0}$ and $\mathbf{1}$ are vectors of zeros and ones with the same dimensions of $\mathbf{b}^{\{l\}}$ and $\boldsymbol{\upsigma}^{\{l\}}$, respectively.

\subsection{$K$-means Clustering}

For shallow RBF-based neural networks, a more sophisticated approach of initialization relies on a clustering algorithm, such as $K$-means, to find~$K=I^{\{1\}}$ cluster centers. Then, these cluster centers as the initial center vectors of RBFs ensure that the centers are distributed along the dataset's inherent structure. However, as the PT-RBF Gaussian neurons operate with a split-complex design, the $K$-means must be applied for the real and imaginary components of the input dataset $\mathcal{X}\supset\mathbf{x}$, separately, creating a set of cluster centers $\mathbf{C}_\mathcal{X}=\mathbf{C}_{\Re{\left(\mathbf{\mathcal{X}}\right)}}+\jmath\mathbf{C}_{\Im{\left(\mathbf{\mathcal{X}}\right)}}$. The $m$-th center vector of the hidden layer is $\boldsymbol{\upgamma}^{\{1\}}_m\in \mathbf{C}_\mathcal{X}$, randomly selected without replacement. 

The center variances are chosen based on the in-cluster distances from $K$-means. Thus, the PT-RBF $m$-th center variance of the hidden layer is
\begin{multline}
    \sigma^{\{1\}}_m = \frac{1}{\left\vert \mathcal{X}_{\Re\left(\boldsymbol{\upgamma}^{\{1\}}_m\right) }\right\vert}\sum_{\Re\left(\mathbf{x}\right) \in \mathcal{X}_{\Re\left(\boldsymbol{\upgamma}^{\{1\}}_m\right) }} \left\Vert \Re\left(\mathbf{x}\right) - \Re\left(\boldsymbol{\upgamma}^{\{1\}}_m\right) \right\Vert_2^2\\
    +\jmath \frac{1}{\left\vert \mathcal{X}_{\Im\left(\boldsymbol{\upgamma}^{\{1\}}_m\right) }\right\vert}\sum_{\Im\left(\mathbf{x}\right) \in \mathcal{X}_{\Im\left(\boldsymbol{\upgamma}^{\{1\}}_m\right) }} \left\Vert \Im\left(\mathbf{x}\right) - \Im\left(\boldsymbol{\upgamma}^{\{1\}}_m\right) \right\Vert_2^2,
\end{multline}
in which $\mathcal{X}_{\Re\left(\boldsymbol{\upgamma}^{\{1\}}_m\right) }\subset \Re\left(\mathcal{X}\right)$ and $\mathcal{X}_{\Im\left(\boldsymbol{\upgamma}^{\{1\}}_m\right) }\subset \Im\left(\mathcal{X}\right)$ are subsets of the input dataset vectors nearest to $\Re\left(\boldsymbol{\upgamma}^{\{1\}}_m\right)$ and $\Im\left(\boldsymbol{\upgamma}^{\{1\}}_m\right)$, respectively. The operator $\vert \cdot\vert$ returns the subset cardinality.

The synaptic weights and bias initializations are equal to the random initialization scheme.

\subsection{Constellation-based initialization}

As an alternative in finite alphabet outputs, the center vectors can be randomly selected from the output dataset~\cite{Loss2007}. For example, when the output dataset is a constellation containing $M$-ary quadrature amplitude modulation~($M$-QAM), all center vector components are initialized with randomly selected $M$-QAM symbols. The PT-RBF $m$-th center variance of the $l$-th hidden layer is
\begin{multline}
    \sigma^{\{l\}}_m = \frac{1}{2}\max_{1\leq i,k\leq I^{\{l\}}}\left[\left\Vert \Re\left(\boldsymbol{\upgamma}^{\{l\}}_i\right) - \Re\left(\boldsymbol{\upgamma}^{\{l\}}_k\right) \right\Vert_2\right]\\
    +\jmath \frac{1}{2}\max_{1\leq i,k\leq I^{\{l\}}}\left[\left\Vert \Im\left(\boldsymbol{\upgamma}^{\{l\}}_i\right) - \Im\left(\boldsymbol{\upgamma}^{\{l\}}_k\right) \right\Vert_2\right].
\end{multline}

The synaptic weights and bias are initialized with zeros.

\section{Deep PT-RBF parameter initialization}
\label{sec:model}


To properly initialize the deep PT-RBF parameters, we first need to understand the relationship between the input vector $\mathbf{x}$ and the Gaussian center vectors $\boldsymbol{\upgamma}_m^{\{1\}}$. In \eqref{eq:kernel}, regarding \eqref{eq:kernel_argument}, and keeping $\sigma_m^{\{1\}}$ constant, the closer $\boldsymbol{\upgamma}_m^{\{1\}}$ is to $\mathbf{x}$, the higher is the value of the real and imaginary parts of $\phi_m^{\{1\}}$. For example, if $\boldsymbol{\upgamma}_m^{\{1\}}=\mathbf{x}$ then $\phi_m^{\{1\}} = 1+\jmath1$. On the other hand, if $\boldsymbol{\upgamma}_m^{\{1\}}$ is set far from $\mathbf{x}$, then $\phi_m^{\{1\}}\to 0$. In this context, in order to not saturate or vanish $\phi_m^{\{1\}}$, we assume $\mu_\mathbf{\bar{x}}=\mu_{\boldsymbol{\upgamma}^{\{1\}}} = 0$ and $\sigma^2_\mathbf{\bar{x}}=\sigma^2_{\boldsymbol{\upgamma}^{\{1\}}}$, where $\mathbf{\bar{x}}$ is the normalized input dataset. Furthermore, we expect that depending on the dataset inputs, $\phi_m^{\{1\}}$ varies reasonably. For example, considering $v_m^{\{1\}}=5$ and $v_m^{\{1\}}=10$, the variation in $\phi_m^{\{1\}}$ is only $4.54\times10^{-5}$. In contrast, considering $v_m^{\{1\}}=0$ and $v_m^{\{1\}}=3$, the variation in $\phi_m^{\{1\}}$ is $0.95$. Then, it is desirable that $\mu_{v^{\{1\}}}$ is not too large. Based on Appendix \ref{app:1}, the expected value of $v^{\{1\}}$ is
\begin{equation}
\label{eq:mean_v_1}
   \mu_{\mathbf{v}^{\{1\}}} = \frac{P}{c_\sigma}\left[\sigma^2_{\Re\left(\mathbf{\bar{x}}\right)}+\jmath\sigma^2_{\Im\left(\mathbf{\bar{x}}\right)}
   +\sigma^2_{\Re\left(\boldsymbol{\upgamma}_m^{\{1\}}\right)}+\jmath\sigma^2_{\Im\left(\boldsymbol{\upgamma}_m^{\{1\}}\right)}\right],
\end{equation}
in which $\sigma^2_{\mathbf{\bar{x}}}=2\sigma^2_{\Re\left(\mathbf{\bar{x}}\right)}=2\sigma^2_{\Im\left(\mathbf{\bar{x}}\right)}$ is the variance of $\mathbf{\bar{x}}$, $\sigma^2_{\boldsymbol{\upgamma}_m^{\{1\}}}=2\sigma^2_{\Re\left(\boldsymbol{\upgamma}_m^{\{1\}}\right)}=2\sigma^2_{\Im\left(\boldsymbol{\upgamma}_m^{\{1\}}\right)}$ is the variance of $\boldsymbol{\upgamma}_m^{\{1\}}$, $c_\sigma=\Re\left(\sigma_m^{\{1\}}\right)=\Im\left(\sigma_m^{\{1\}}\right)\,\,\forall \,\,m$, and $c_\sigma$ is a positive and nonzero constant.

As $\sigma^2_\mathbf{\bar{x}}=\sigma^2_{\boldsymbol{\upgamma}^{\{1\}}}$, from \eqref{eq:mean_v_1}, we have
\begin{equation}
\label{eq:var_x}
    \sigma^2_\mathbf{\bar{x}} = \frac{c_\sigma\mu_{\Re\left(\mathbf{v}^{\{1\}}\right)}}{P}.
\end{equation}

Based on \eqref{eq:var_x}, the normalized input is given as
\begin{equation}
\label{eq:norm_x}
    \mathbf{\bar{x}} = \frac{(\mathbf{x}-\mu_{\mathbf{x}})}{\sqrt{\sigma^2_{\mathbf{x}}}} \sqrt{\frac{c_\sigma \mu_{\Re\left(\mathbf{v}^{\{1\}}\right)}}{P}},
\end{equation}
where $\mu_{\mathbf{x}}$ and $\sigma^2_{\mathbf{x}}$ are applied to adjust the mean and variance of $\mathbf{\bar{x}}$ before the normalization by \eqref{eq:var_x}.

Similarly, in the first hidden layer, the normalized matrix of center vectors is
\begin{equation}
\label{eq:gamma_init}
    \boldsymbol{\Gamma}^{\{1\}} \sim \mathcal{C}\mathcal{G}\left(0,\frac{c_\sigma \mu_{\Re\left(\mathbf{v}^{\{1\}}\right)}}{P} \right).
\end{equation}

In order to normalize the output dataset $\mathbf{d}$, we need to compute the variance of the output vector $\mathbf{y}^{\{L\}}$, by
\begin{equation}
\label{eq:variance_y_output}
    \sigma^2_{\mathbf{y}^{\{L\}}} = \mathrm{Var}\left[\mathbf{W}^{\{L\}}\boldsymbol{\upphi}^{\{L\}}+\mathbf{b}^{\{L\}}\right].
\end{equation}

However, as $\mathbf{W}^{\{L\}}$ is a complex-valued matrix and performs a linear combination with $\boldsymbol{\upphi}^{\{L\}}$, which is a complex-valued vector, the real and imaginary components are handled at the same time, in the complex domain. Moreover, we assume that $\mathbf{b}^{\{l\}}$ is initialized with zeros, for all layers. Thus, based on Appendix~\ref{app:3}, \eqref{eq:variance_y_output} results in
\begin{equation}
\sigma _{\mathbf{y}^{\{L\}}}^{2} =\frac{12}{5} c_{\sigma }^{-2}\exp (-2\mu _{\mathbf{v}^{\{L\}}} )I^{\{L\}}\mathrm{O^{\{L-1\}}} \sigma _{\mathbf{W}^{\{L\}}}^{2} \sigma _{\upgamma ^{\{L\}}}^{4},
\end{equation}
where $\sigma^2_{\mathbf{W^{\{L\}}}}$ is the variance of $\mathbf{W}^{\{L\}}$, and $\mu_{\mathbf{v}^{\{L\}}}$ is the expected value of $\mathbf{v}^{\{L\}}$. Choosing $\sigma^2_{\mathbf{y}^{\{L\}}}=\sigma^2_\mathbf{\bar{d}}$, i.e., the variance of the PT-RBF output equal to the variance of the normalized output dataset, yields the initialization of $\mathbf{W}^{\{L\}}$ as
\begin{equation}
\label{eq:w_init}
    \mathbf{W}^{\{L\}} \sim \mathcal{C}\mathcal{G}\left(0,\frac{5 \exp (2\mu _{\mathbf{v}^{\{L\}}} ) \sigma^2_{\mathbf{\bar{d}}}}{12 c_{\sigma }^{-2}I^{\{L\}}\mathrm{O^{\{L-1\}}} \sigma _{\mathbf{W}^{\{L\}}}^{2} \sigma _{\upgamma ^{\{L\}}}^{4}} \right),
\end{equation}
in which the output dataset can be normalized by
\begin{equation}
\label{eq:norm_d}
    \mathbf{\bar{d}} = \frac{(\mathbf{d}-\mu_{\mathbf{d}})}{\sqrt{\sigma^2_{\mathbf{d}}}} \sqrt{\sigma^2_{\mathbf{y}^{\{L\}}}}=\frac{(\mathbf{d}-\mu_{\mathbf{d}})}{\sqrt{\sigma^2_{\mathbf{d}}}} \sqrt{\frac{c_\sigma\mu_{\Re\left(\mathbf{v}^{\{L\}}\right)}}{R}}.
\end{equation}

Relying on \eqref{eq:gamma_init}, we can generalize the initialization of $\boldsymbol{\Gamma}^{\{l\}}$, as
\begin{equation}
\label{eq:gamma_init_layers}
    \boldsymbol{\Gamma}^{\{l\}} \sim \mathcal{C}\mathcal{G}\left(0,\frac{c_\sigma \mu_{\Re\left(\mathbf{v}^{\{l\}}\right)}}{O^{\{l-1\}}} \right).
\end{equation}

From \eqref{eq:var_x}, the variance of the output hidden layers can be considered as
\begin{equation}
\label{eq:var_y_hidden}
    \sigma^2_{\mathbf{y}^{\{l\}}} = \frac{c_\sigma\mu_{\Re\left(\mathbf{v}^{\{l+1\}}\right)}}{O^{\{l\}}},
\end{equation}
where, replacing  $\sigma _{\upgamma ^{\{L\}}}^{4}$ by \eqref{eq:gamma_init_layers} and $\sigma^2_{\mathbf{\bar{d}}}$ by \eqref{eq:var_y_hidden} into \eqref{eq:w_init}, yields
\begin{equation}
\label{eq:w_init_layers}
    \mathbf{W}^{\{l\}} \sim \mathcal{CG}\left( 0,\frac{5c_{\sigma } \exp (2\mu _{\mathbf{v}^{\{L\}}} ) O^{\{l-1\}}}{12{I^{\{l\}}} O^{\{l\}} \mu _{ \mathbf{v}^{\{l\}})}}\right),
\end{equation}
It is important to note that, \eqref{eq:norm_x} and \eqref{eq:norm_d} only hold for $\sigma^2_{\mathbf{x}}=2\sigma^2_{\Re\left(\mathbf{x}\right)}=2\sigma^2_{\Im\left(\mathbf{x}\right)}$ and $\sigma^2_{\mathbf{d}}=2\sigma^2_{\Re\left(\mathbf{d}\right)}=2\sigma^2_{\Im\left(\mathbf{d}\right)}$, respectively. For the particular case of $\sigma^2_{\Re\left(\mathbf{x}\right)}\neq\sigma^2_{\Im\left(\mathbf{x}\right)}$ and $\sigma^2_{\Re\left(\mathbf{d}\right)}\neq\sigma^2_{\Im\left(\mathbf{d}\right)}$, then \eqref{eq:norm_x} and \eqref{eq:norm_d} become
\begin{multline}
    \mathbf{\bar{x}} = \left[\frac{(\Re\left(\mathbf{x}\right)-\mu_{\Re\left(\mathbf{x}\right)})}{\sqrt{2\sigma^2_{\Re\left(\mathbf{x}\right)}}}+\jmath\frac{(\Im\left(\mathbf{x}\right)-\mu_{\Im\left(\mathbf{x}\right)})}{\sqrt{2\sigma^2_{\Im\left(\mathbf{x}\right)}}}\right] \\
    \times\sqrt{\frac{c_\sigma \mu_{\Re\left(\mathbf{v}^{\{1\}}\right)}}{P}},
\end{multline}
\begin{multline}
    \mathbf{\bar{d}} = \left[\frac{(\Re\left(\mathbf{d}\right)-\mu_{\Re\left(\mathbf{d}\right)})}{\sqrt{2\sigma^2_{\Re\left(\mathbf{d}\right)}}}+\jmath\frac{(\Im\left(\mathbf{d}\right)-\mu_{\Im\left(\mathbf{d}\right)})}{\sqrt{2\sigma^2_{\Im\left(\mathbf{d}\right)}}}\right] \\\times\sqrt{\frac{c_\sigma\mu_{\Re\left(\mathbf{v}^{\{L\}}\right)}}{R}}.
\end{multline}

\section{Results}
\label{sec:results}


For the sake of simplification, in the proposed approach, the parameters $c_\sigma$ and $\mu_{\mathbf{v}^{\{l\}}}$ were set to $1$, for all layers. Then, the initializations and normalizations become
\begin{equation}
    \mathbf{b}^{\{l\}}=\mathbf{0}+\jmath\mathbf{0},
\end{equation}
\begin{equation}
    \boldsymbol{\upsigma}^{\{l\}}=\mathbf{1}+\jmath\mathbf{1},
\end{equation}
\begin{equation}
    \boldsymbol{\Gamma}^{\{l\}} \sim \mathcal{C}\mathcal{G}\left(0,\frac{1}{O^{\{l-1\}}} \right),
\end{equation}
\begin{equation}
    \mathbf{W}^{\{l\}} \sim \mathcal{CG}\left( 0,\frac{5\exp (2)O^{\{l-1\}}}{12I^{\{l\}} O^{\{l\}}}\right),
\end{equation}
\begin{equation}
    \mathbf{\bar{x}} = \frac{(\mathbf{x}-\mu_{\mathbf{x}})}{\sqrt{\sigma^2_{\mathbf{x}}}} \sqrt{\frac{1}{P}},
\end{equation}
\begin{equation}
    \mathbf{\bar{d}} = \frac{(\mathbf{d}-\mu_{\mathbf{d}})}{\sqrt{\sigma^2_{\mathbf{d}}}} \sqrt{\frac{1}{R}}.
\end{equation}

For the random initialization, we defined $\sigma^2_{\boldsymbol{\Gamma}^{\{l\}}}=1$. The $K$-means and constellation-based initializations are obtained from the input and output datasets, respectively.

Based on \cite{Soares2023}, we consider a space-time block coding~(STBC) simulation system with the 3GPP TS 38.211 specification for 5G physical channels and modulation~\cite{3gpp.38.211}. The orthogonal frequency-division multiplexing~~(OFDM) is defined with 60-kHz subcarrier spacing, 256 active subcarriers, and a block-based pilot scheme. Symbols are modulated with~16-QAM and, for the MIMO setup, 4~antennas are employed both at the transmitter and receiver. Based on the tapped delay line-A (TDLA) from the 3GPP TR 38.901 5G channel models~\cite{3gpp.38.901}, the MIMO channel follows the TDLA from the 3GPP TR 38.104 5G radio base station transmission and reception~\cite{3gpp.38.104}. The TDLA is described with 12~taps, with varying delays from 0.0~ns to 290~ns and powers from -26.2~dB to 0~dB. A Rayleigh distribution is used to compute each sub-channel. To avoid influencing the learning curves, we do not take into account the Doppler effect, and we do not employ the inference learning techniques proposed in~\cite{Soares2023}. The CVNNs operate with 16 inputs and 4 outputs. The inputs are taken from the OFDM demodulator outputs, one at a time~(see \cite{Soares2023}, Fig. 1). Training and validation were performed for $3,840$ and $1,280$ instances, respectively. To assess performance, we calculated the Mean Squared Error (MSE), defined as $\text{MSE}=\frac{1}{n} \sum_{i=1}^{n} (y_i - \hat{y}_i)^2$, where $y_i$ represents the total transmitted constellation symbols over 100 simulations. Each output training instance corresponds to 4 symbols, resulting in a total of 15,360 and 5,120 16-QAM symbols for training and validation phases, respectively.

\begin{figure}[t]
    \centering    \includegraphics[width=1\linewidth]{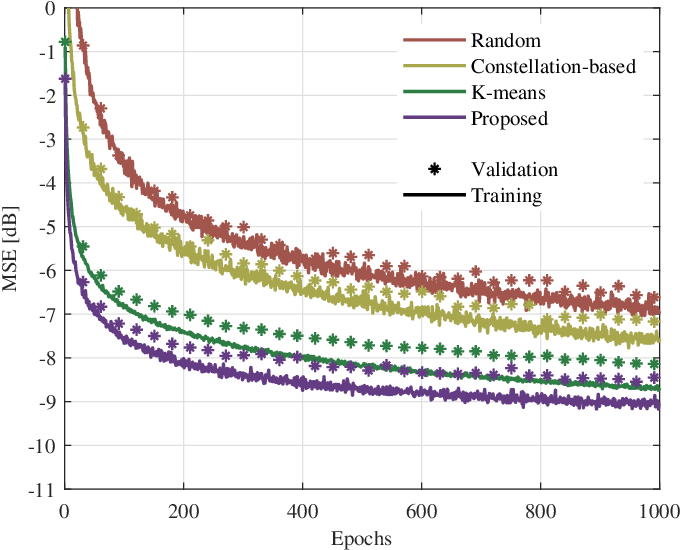}
    \caption{MSE convergence results of training~(solid lines) and validation~(asterisks) of the PT-RBF initialization with a hidden layer~($I^{\{1\}}=64$~neurons) for joint channel estimation and decoding in a MIMO-OFDM $4\times4$ system, operating with 16-QAM and 256 subcarriers. Results were averaged over 100~subsequent simulations with $\mathrm{E_b/N_0}=26$~dB.}
    \label{fig:MIMO44_results1L}
\end{figure}

Fig.~\ref{fig:MIMO44_results1L} illustrates the MSE convergence for 1000~epochs of training~(solid lines) and validation~(asterisks) of the PT-RBF with a hidden layer~($I^{\{1\}}=64$~neurons). Results were averaged over $100$~simulations with a bit energy to noise power spectral density ratio~$\mathrm{E_b/N_0}=26$~dB. Table~\ref{tab:parameters_single} depicts the PT-RBF hyperparameters empirically optimized for each initialization scheme. None of the algorithms presented under- or over-fitting. The random initialization presented the poorest convergence results, with a steady-state error of $-6$~dB. On the other hand, the constellation-based and $K$-means initializations achieved a steady-state error of $-6$~dB and $-8.8$~dB, respectively. The best results were obtained with the proposed approach, with $-9.1$~dB of steady-state error. For comparison results regarding the convergence rate, considering $\mathrm{MSE}=-5$~dB, the proposed approach reaches this mark in five training epochs, followed by the $K$-means~(11~epochs), constellations-based~(80~epochs), and random initialization~(165~epochs).
\begin{table}[t]
\centering
\renewcommand{\arraystretch}{1.3}
\begin{threeparttable}
\caption{Single layer PT-RBF optimized hyperparameters.}
\label{tab:parameters_single}
\begin{tabular}{l c c c c}
\hline
Algorithm &  $\eta_w$ & $\eta_b$ & $\eta_\gamma$ & $\eta_\sigma$\\
\hline
Random  & 0.5 & 0.5 & 0.5 & 0.5\\
Constellation-based  & 0.5 & 0.5 & 0.5 & 0.5\\
$K$-means  & 0.1 & 0.1 & 0.4 & 0.2\\
Proposed Approach & 0.1 & 0.1 & 0.4 & 0.2\\
\hline
\end{tabular}
\end{threeparttable}
\end{table}

\begin{figure}[t]
    \centering
\includegraphics[width=1\linewidth]{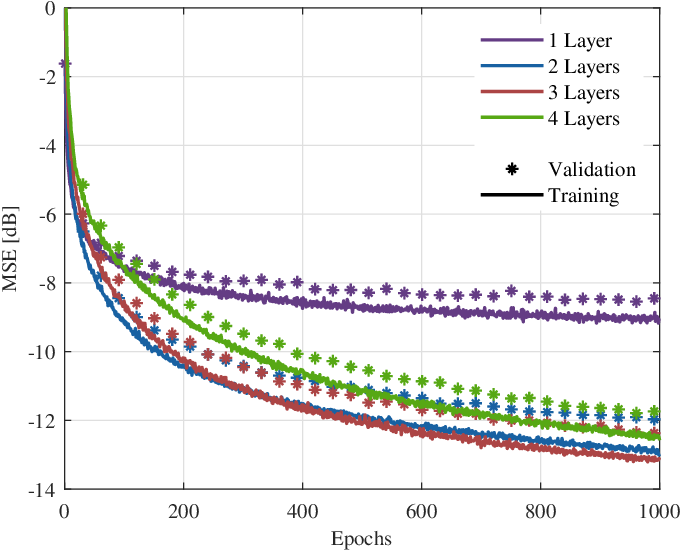}
    \caption{MSE convergence results of training~(solid lines) and validation~(stars) of the proposed initialization approach with one~($I^{\{1\}}=64$~neurons), two~($I^{\{1\}}=48$ and $I^{\{2\}}=16$~neurons), three~($I^{\{1\}}=24$, $I^{\{2\}}=24$, and $I^{\{3\}}=16$~neurons), and four~($I^{\{1\}}=16$, $I^{\{2\}}=16$, $I^{\{3\}}=16$, and $I^{\{4\}}=16$~neurons) hidden layers for joint channel estimation and decoding in a MIMO-OFDM $4\times4$ system, operating with 16-QAM and 256 subcarriers. Results were averaged over 100 subsequent simulations with $\mathrm{E_b/N_0}=26$~dB.}
    \label{fig:MIMO44_results2L}
\end{figure}

\begin{table}[t]
\centering
\renewcommand{\arraystretch}{1.3}
\begin{threeparttable}
\caption{Deep PT-RBF optimized hyperparameters for the proposed approach.}
\label{tab:parameters_deep}
\begin{tabular}{l c c c c}
\hline
Algorithm &  $\eta_w$ & $\eta_b$ & $\eta_\gamma$ & $\eta_\sigma$\\
\hline
first layer  & 0.100 & 0.100 & 0.100 & 0.100\\
second layer & 0.050 & 0.050 & 0.050 & 0.050\\
third layer  & 0.033 & 0.033 & 0.033 & 0.033\\
fourth layer & 0.025 & 0.025 & 0.025 & 0.025\\
\hline
\end{tabular}
\begin{tablenotes}
\item These hyperparameters were optimized for the proposed initialization of the deep PT-RBFs. For example, in a deep PT-RBF with two hidden layers, only the first and second rows of hyperparameters are necessary. In a shallow architecture, the optimization is available in Table~\ref{tab:parameters_single}.
\end{tablenotes}
\end{threeparttable}
\end{table}

For further comparison, we have also employed the initialization schemes for PT-RBFs with two, three, and four hidden layers. However, the $K$-means was not considered since it is only suitable for shallow RBFs. In addition, although several trials were attempted, no convergence was achieved for the random and constellation-based initializations. Thus, Fig.~\ref{fig:MIMO44_results2L} shows the convergence results for the proposed approach for the PT-RBFs with one~($I^{\{1\}}=64$~neurons), two~($I^{\{1\}}=48$ and $I^{\{2\}}=16$~neurons), three~($I^{\{1\}}=24$, $I^{\{2\}}=24$, and $I^{\{3\}}=16$~neurons), and four~($I^{\{1\}}=16$, $I^{\{2\}}=16$, $I^{\{3\}}=16$, and $I^{\{4\}}=16$~neurons) hidden layers\footnote{For the sake of comparison, we chose a total number of neurons $N_T=64$, which was split depending on the number of layers.}. Table~\ref{tab:parameters_deep} depicts the deep PT-RBF hyperparameters empirically optimized for each hidden layer. Unlike the other initialization schemes, the proposed approach achieves reasonable convergence for all architectures. One may note that all multilayered PT-RBFs architectures achieved the same steady-state MSE results. This result is due to the number of neurons utilized to create the PT-RBF layers. For the three- and four-layered PT-RBFs, the layers with the lowest number of neurons performed bottlenecks, impacting results. In order to circumvent this issue, more neurons could be adopted per layer; nonetheless, it does not affect the convergence verification.

\section{Conclusion}
\label{sec:conclusion}

This paper presents an in-depth analysis of the initialization process in phase-transmittance radial basis function~(PT-RBF) neural networks. Our findings have elucidated the intricate dependencies involved in the initialization process. Specifically, the normalization between layers which is dependent on the number of inputs, outputs and neurons. This reveals that synaptic weights initialization is influenced by the layer-wise configuration of inputs, neurons, and outputs. Consequently, our proposed approach demonstrates robustness to variations in the number of inputs, outputs, hidden layers, and neurons.

This innovation is particularly impactful for deploying these networks in real-world scenarios, which require robustness for a wide range of different configurations, with no room for ad hoc adjustments. In a carefully designed simulation environment, our proposed deep PT-RBF parameter initialization exhibited superior convergence performance compared to existing methods such as random, $K$-means, and constellation-based initialization. Notably, for multi-layer architecture, our method was the only one that achieved successful convergence, highlighting its unique efficacy and adaptability. 

The results of our study have important implications. Firstly, they introduce a robust and effective initialization method that can significantly improve the convergence rate and steady state MSE of PT-RBF neural networks. Additionally, offering the potential for extending our framework to other RBF neural network architectures. In future works, we plan to validate the robustness of our proposed approach through more exhaustive experiments. We also aim to explore the applicability of our initialization framework to other neural network architectures, thereby contributing to the broader advancement of neural network-based solutions in digital communications.

\section*{Acknowledgments}

This study was supported in part by the Coordenação de Aperfeiçoamento de Pessoal de Nível Superior --- Brasil
(CAPES) --- Finance Code 001.

\appendices

\section{Expected value of $\mathbf{v}^{\{l\}}$}
\label{app:1}

Taking the Gaussian kernel input $\mathbf{v}^{\{l\}}$ of a layer $l$, its expected value is
\begin{multline}
\label{eq:mean_v_l}
 \mu_{\mathbf{v}^{\{l\}}} = \mathrm{E}\left[\vphantom{\frac{\left \Vert\Re\left(\mathbf{y}^{\{l\}}\right)-\Re\left(\boldsymbol{\upgamma}_m^{\{l\}}\right)\right \Vert_2^2}{\Re\left(\sigma_m^{\{l\}}\right)}}\right.\frac{\left \Vert\Re\left(\mathbf{y}^{\{l-1\}}\right)-\Re\left(\boldsymbol{\upgamma}_m^{\{l\}}\right)\right \Vert_2^2}{\Re\left(\sigma_m^{\{l\}}\right)} \\
+\jmath\frac{\left \Vert\Im\left(\mathbf{y}^{\{l-1\}}\right)-\Im\left(\boldsymbol{\upgamma}_m^{\{l\}}\right)\right \Vert_2^2}{\Im\left(\sigma_m^{\{l\}}\right)}\left.\vphantom{\frac{\left \Vert\Re\left(\mathbf{y}^{\{l\}}\right)-\Re\left(\boldsymbol{\upgamma}_m^{\{l\}}\right)\right \Vert_2^2}{\Re\left(\sigma_m^{\{l\}}\right)}}\right],
\end{multline}
where $m\in[1,\,2,\cdots,I^{\{l\}}]$.

Due to the split-complex kernel of PT-RBFs, the expected value of the real and imaginary components can be computed separately. Focusing on the real part
\begin{equation}
    \Re\left(\mu_{\mathbf{v}^{\{l\}}}\right) = \mathrm{E}\left[\frac{\left \Vert\Re\left(\mathbf{y}^{\{l-1\}}\right)-\Re\left(\boldsymbol{\upgamma}_m^{\{l\}}\right)\right \Vert_2^2}{\Re\left(\sigma_m^{\{l\}}\right)}\right].
\end{equation}

Assuming $\Re\left(\sigma_m^{\{l\}}\right)=\Im\left(\sigma_m^{\{l\}}\right)=c_\sigma \,\,\, \forall m, \{l\}$, with $c_\sigma > 0$, then
\begin{equation}
    \Re\left(\mu_{\mathbf{v}^{\{l\}}}\right) = c_\sigma^{-1}\mathrm{E}\left[\left \Vert\Re\left(\mathbf{y}^{\{l-1\}}\right)-\Re\left(\boldsymbol{\upgamma}_m^{\{l\}}\right)\right \Vert_2^2\right],
\end{equation}
which is
\begin{multline}
    \Re\left(\mu_{\mathbf{v}^{\{l\}}}\right) = c_\sigma^{-1}\sum_{i=1}^{O^{\{l-1\}}}\mathrm{E}\left[\Re\left(y_i^{\{l-1\}}\right)^2\right]+\mathrm{E}\left[\Re\left(\gamma_{m,i}^{\{l\}}\right)^2\right]\\-2\mathrm{E}\left[\Re\left(y_i^{\{l-1\}}\right)\Re\left(\gamma_{m,i}^{\{l\}}\right)\right],
\end{multline}
where $\gamma_{m,i}^{\{l\}}$ and $y_i^{\{l-1\}}$ are the $i$-th elements of $\boldsymbol{\upgamma}_m^{\{l\}}$ and $\mathbf{y}^{\{l-1\}}$, respectively.

As $\boldsymbol{\upgamma}_m^{\{l\}}$ and $\mathbf{y}^{\{l-1\}}$ are independent, and assuming $\boldsymbol{\upgamma}_m^{\{l\}}$ and $\mathbf{y}^{\{l-1\}}$ with zeros mean, then
\begin{equation}
\label{eq:mean_v_zero}
\mathrm{E}\left[\Re\left(y_i^{\{l-1\}}\right)\Re\left(\gamma_{m,i}^{\{l\}}\right)\right]=0,
\end{equation}
which results in
\begin{equation}
\label{eq:mu_v_r_part1}
   \Re\left(\mu_{\mathbf{v}^{\{l\}}}\right) = c_\sigma^{-1}\sum_{i=1}^{O^{\{l-1\}}}\mathrm{E}\left[\Re\left(y_i^{\{l-1\}}\right)^2\right]+\mathrm{E}\left[\Re\left(\gamma_{m,i}^{\{l\}}\right)^2\right].
\end{equation}

Furthermore, the variances of $\Re\left(\mathbf{y}^{\{l-1\}}\right)$ and $\Re\left(\boldsymbol{\upgamma}_m^{\{l\}}\right)$ are
\begin{equation}
\label{eq:sigma_y}
\sigma^2_{\Re\left(\mathbf{y}^{\{l-1\}}\right)} = \frac{1}{O^{\{l-1\}}}\sum_{i=1}^{O^{\{l-1\}}} \Re\left(y_i^{\{l-1\}}\right)^2,
\end{equation}
\begin{equation}
\label{eq:sigma_gamma}
\sigma^2_{\Re\left(\boldsymbol{\upgamma}_m^{\{l\}}\right)} = \frac{1}{O^{\{l-1\}}}\sum_{i=1}^{O^{\{l-1\}}} \Re\left(\gamma_i^{\{l\}}\right)^2,
\end{equation}
thus, in \eqref{eq:mu_v_r_part1}, applying the summation to the expected value arguments, yields
\begin{equation}
   \Re\left(\mu_{\mathbf{v}^{\{l\}}}\right) = c_\sigma^{-1}O^{\{l-1\}}\mathrm{E}\left[ \sigma^2_{\Re\left(\mathbf{y}^{\{l-1\}}\right)}+\sigma^2_{\Re\left(\boldsymbol{\upgamma}_m^{\{l\}}\right)}\right].
\end{equation}

However, as the variances $\sigma^2_{\Re\left(\mathbf{y}^{\{l-1\}}\right)}$ and $\sigma^2_{\Re\left(\boldsymbol{\upgamma}_m^{\{l\}}\right)}$ are constants, $\Re\left(\mu_{\mathbf{v}^{\{l\}}}\right)$ can be expressed as 
\begin{equation}
   \Re\left(\mu_{\mathbf{v}^{\{l\}}}\right) = c_\sigma^{-1}O^{\{l-1\}}\left[ \sigma^2_{\Re\left(\mathbf{y}^{\{l-1\}}\right)}+\sigma^2_{\Re\left(\boldsymbol{\upgamma}_m^{\{l\}}\right)}\right].
\end{equation}

Applying these computations to $\Re\left(\mu_{\mathbf{v}^{\{l\}}}\right)$, and adding it to $\Re\left(\mu_{\mathbf{v}^{\{l\}}}\right)$, we obtain the expected value of $\mathbf{v}^{\{l\}}$ as 
\begin{multline}
\label{eq:mean_v_final}
   \mu_{\mathbf{v}^{\{l\}}} = c_\sigma^{-1}O^{\{l-1\}}\left[\vphantom{\sigma^2_{\Re\left(\boldsymbol{\upgamma}_m^{\{l\}}\right)}+\jmath\sigma^2_{\Im\left(\boldsymbol{\upgamma}_m^{\{l\}}\right)}}\sigma^2_{\Re\left(\mathbf{y}^{\{l-1\}}\right)}+\jmath\sigma^2_{\Im\left(\mathbf{y}^{\{l-1\}}\right)}\right.\\
   + \left.\sigma^2_{\Re\left(\boldsymbol{\upgamma}_m^{\{l\}}\right)}+\jmath\sigma^2_{\Im\left(\boldsymbol{\upgamma}_m^{\{l\}}\right)}\right].
\end{multline}

\section{Variance of $\mathbf{v}^{\{l\}}$}
\label{app:2}

The variance of $\mathbf{v}^{\{l\}}$ is
\begin{equation}
\label{eq:var_v_init}
\sigma _{\mathbf{v}^{\{l\}}}^{2} =\text{Var}\left[ \Re \left(\mathbf{v}^{\{l\}}\right)\right] +\text{Var}\left[ \Im \left(\mathbf{v}^{\{l\}}\right)\right],
\end{equation}
because the PT-RBF split-complex kernel.

First, considering the term $\text{Var}\left[ \Re\left(\mathbf{v}^{\{l\}}\right) \right]$, we have
\begin{multline}
\label{eq:var_v_init_open}
    \text{Var}\left[ \Re \left(\mathbf{v}^{\{l\}}\right)\right] = \left(\frac{O^{\{l-1\}}}{c_{\sigma }}\right)^{2} \text{Var}\left[\mathrm{E}\left[ \Re \left( y_{i}^{\{l-1\}}\right)^{2}\right] \right. \\
    \left.+\mathrm{E}\left[ \Re \left( {\upgamma}_m^{\{l\}}\right)^{2}\right] -\mathrm{E}\left[ 2\Re \left( y_{i}^{\{l-1\}} {\upgamma}_m^{\{l\}}\right)\right]\right].
\end{multline}

Theoretically, $\text{Var}\left[\mathrm{E}\left[ \Re \left( {\upgamma}_m^{\{l\}}\right)^{2}\right] -\mathrm{E}\left[ 2\Re \left( y_{i}^{\{l-1\}} {\upgamma}_m^{\{l\}}\right)\right]\right]$ should be zero, since $\mathrm{E}\left[ \Re \left( {\upgamma}_m^{\{l\}}\right)^{2}\right]$ is the variance of  ${\upgamma}_m^{\{l\}}$ and for $\mathrm{E}\left[ 2\Re \left( y_{i}^{\{l-1\}} {\upgamma}_m^{\{l\}}\right)\right]$ both ${\upgamma}_m^{\{l\}}$ and $ y_{i}^{\{l-1\}} $ have zero mean.
However, in practice, as our vector is limited in size, we will have an approximation with a small value.
In fact, $\text{Var}\left[\mathrm{E}\left[ 2\Re \left( y_{i}^{\{l-1\}} {\upgamma}_m^{\{l\}}\right)\right]\right]$ is the variance of the sample mean, which is given by $\sigma^2 (\mu _{\overline{x}} )=\frac{\sigma _{x}^{2}}{n}$. Thus, considering the variance of $y_{i}^{\{l-1\}} {\upgamma}_m^{\{l\}}$ as the product of the variances of $y_{i}^{\{l-1\}}$ and ${\upgamma}_m^{\{l\}} $, denoted as $\sigma _{y_{i}^{\{l-1\}}}^{2}$ and $\sigma _{{\upgamma}_m^{\{l\}}}^{2}$ respectively, which is valid since they are independent, we have
\begin{equation}
\label{eq:var_v_var_mean}
    \text{Var}\left[\mathrm{E}\left[ 2\Re \left( y_{i}^{\{l-1\}} \upgamma _{m}^{\{l\}}\right)\right]\right] =\frac{4 \Re \left( \sigma _{y_{i}^{\{l-1\}}}^{2} \sigma _{\upgamma _{m}^{\{l\}}}^{2}\right)}{{O^{\{l-1\}}}},
\end{equation}
and similarly $\text{Var}\left[\mathrm{E}\left[ \Re \left( {\upgamma}_m^{\{l\}}\right)^{2}\right]\right]$ is the variance of the sample variance given by $\displaystyle \text{Var}\left(\hat{\sigma }^{2}\right) =\frac{\mu _{4} -\frac{n-3}{n-1} \sigma ^{4}}{n}$, where $\displaystyle \mu _{4}$ is the fourth central moment, given by $\displaystyle \mu _{4} =\frac{\left(\sqrt{12\sigma ^{2}}\right)^{4}}{80}$for $\mathcal{U}\left( 0,\sigma ^{2}\right)$, then
\begin{multline}
    \text{Var}\left[\mathrm{E}\left[ \Re \left( {\upgamma}_m^{\{l\}}\right)^{2}\right]\right] = \frac{1}{{O^{\{l-1\}}}}\\\times\left(\frac{\left(\sqrt{12 \Re \left( \sigma _{\upgamma _{m}^{\{l\}}}^{2}\right)}\right)^{4}}{80} -\frac{{O^{\{l-1\}}} -3}{{O^{\{l-1\}}} -1} \left(\sqrt{\Re \left( \sigma _{\upgamma _{m}^{\{l\}}}^{2}\right)}\right)^{4}\right),
\end{multline}
which simplifies to
\begin{equation}
\label{eq:var_v_var_var}
    \text{Var}\left[\mathrm{E}\left[ \Re \left( \upgamma _{m}^{\{l\}}\right)^{2}\right]\right] =\frac{4 \Re \left( \sigma _{\upgamma _{m}^{\{l\}}}^{2}\right)^{2}}{5{O^{\{l-1\}}}}.
\end{equation}

In view of $\text{Var}\left[ \Re \left( y_{i}^{\{l-1\}}\right)^{2}\right]$ is a constant in regard to $m$, then $\text{Var}\left[\mathrm{E}\left[ \Re \left( y_{i}^{\{l-1\}}\right)^{2}\right]\right] = 0$, and  $\text{Var}\left[ \Re\left(\mathbf{v}^{\{l\}}\right) \right] = \text{Var}\left[ \Im\left(\mathbf{v}^{\{l\}}\right) \right]$. Substituting \eqref{eq:var_v_var_mean}, and \eqref{eq:var_v_var_var} into \eqref{eq:var_v_init_open}, we obtain
\begin{multline}
    \text{Var}\left[  \left(\mathbf{v}^{\{l\}}\right)\right] = 2 \left(\frac{O^{\{l-1\}}}{c_{\sigma }}\right)^{2}\\ \times\left( \frac{4\Re \left( \sigma _{y_{i}^{\{l-1\}}}^{2} \sigma _{\upgamma _{m}^{\{l\}}}^{2}\right)}{{O^{\{l-1\}}}}
    + \frac{4\left( \Re \left( \sigma _{\upgamma _{m}^{\{l\}}}^{2}\right)\right)^{2}}{5{O^{\{l-1\}}}} \right).
\end{multline}

Since $\sigma^2_\mathbf{\bar{x}} = \sigma _{y_{i}^{\{l-1\}}}^{2} = \sigma^2_{\boldsymbol{\upgamma}_m^{\{1\}}}$, from \eqref{eq:mean_v_1}, we have

\begin{equation}
\label{eq:var_v_final}
   \text{Var}\left[\left(\mathbf{v}^{\{l\}}\right)\right] =\frac{12}{5} c_{\sigma }^{-2}{O^{\{l-1\}}} \sigma _{\upgamma _{m}^{\{l\}}}^{4}.
\end{equation}

\section{Variance of $\mathbf{y}^{\{l\}}$}
\label{app:3}

The variance of $\mathbf{y}^{\{l\}}$ is
\begin{equation}
\label{eq:var_y_init}
    \sigma _{\mathbf{y}^{\{l\}}}^{2} = \text{Var}\left[\left(\mathbf{w}_{m}^{\{l\}}\right)^{T}\boldsymbol{\upphi }^{\{l\}}\right],
\end{equation}
where $m\in[1,\,2,\cdots,I^{\{l\}}]$.
Differently from \eqref{eq:var_v_init}, we cannot compute the expected value of the real and imaginary components separately since $\mathbf{y}^{\{l\}}$ is a linear combination result of a complex-valued vector $\boldsymbol{\upphi}^{\{l\}}$ and a complex-valued matrix $\mathbf{W}^{\{l\}}$.  Expanding  \eqref{eq:var_y_init}
\begin{equation}
    \sigma _{\mathbf{y}^{\{l\}}}^{2} = \text{Var}\left[\sum _{m=1}^{I^{\{l\}}} \phi _{m}^{\{l\}} w_{i,m}^{\{l\}}\right],
\end{equation}
where $\phi _{m}^{\{l\}}$ and $w_{i,m}^{\{l\}}$ are the $m$-th elements of $\boldsymbol{\upphi }^{\{l\}}$ and $\mathbf{w}_{i}^{\{l\}}$ respectively, and $i$ is the $i$-th element of $\mathbf{y}^{\{l\}}$. In addition
\begin{equation}
\label{eq:var_y_fourth}
    \sigma _{\mathbf{y}^{\{l\}}}^{2} = {I{^{\{l\}}}}^{2} \text{Var}\left[\mathrm{E}\left[ \boldsymbol{\upphi } ^{\{l\}} \mathbf{w}_{i}^{\{l\}}\right]\right].
\end{equation}

From \eqref{eq:var_y_fourth} the variance $\sigma _{\mathbf{y}^{\{l\}}}^{2}$ incurs in the variance of the sample mean, since they are independent, we have, $\text{Var}\left[ \boldsymbol{\upphi } ^{\{l\}} \mathbf{w}_{i}^{\{l\}}\right] = \sigma _{\mathbf{w}_{i}^{\{l\}}}^{2} \sigma _{\boldsymbol{\upphi } ^{\{l\}}}^{2}$, then
\begin{equation}
\label{eq:var_mean_phi_w}    \text{Var}\left[\mathrm{E}\left[ \boldsymbol{\upphi } ^{\{l\}} \mathbf{w}_{i}^{\{l\}}\right]\right] = \frac{\sigma _{\mathbf{w}^{\{l\}}}^{2} \sigma _{\boldsymbol{\upphi } ^{\{l\}}}^{2}}{{I^{\{l\}}}}.
\end{equation}

Substituting \eqref{eq:var_mean_phi_w} into \eqref{eq:var_y_fourth}, we obtain the expression
\begin{equation}
\label{eq:var_y_IWphi}
    \sigma _{\mathbf{y}^{\{l\}}}^{2} = \mathrm{I^{\{l\}}} \sigma _{\mathbf{w}^{\{l\}}}^{2} \sigma _{\boldsymbol{\upphi } ^{\{l\}}}^{2}.
\end{equation}

From \eqref{eq:kernel}, and in view of $\Re \left( \mathbf{v}^{{\{l\}}}\right) =\Im \left( \mathbf{v}^{{\{l\}}}\right)$, thus $\sigma _{{\boldsymbol{\upphi }} ^{\{l\}}}^{2} =2\text{Var}\left\{\exp\left[ -\Re \left( \mathbf{v}^{{\{l\}}}\right)\right]\right\}$, 
In order to calculate  $\text{Var}\left\{\exp\left[ -\Re \left( \mathbf{v}^{\{l\}}\right)\right]\right\} \ =\mathrm{E}\left\{\exp\left[ -\Re \left( \mathbf{v}^{{\{l\}}} \right)\right]^{2}\right\} -\mathrm{E}\left\{\exp\left[ -\Re \left( \mathbf{v}^{{\{l\}}}\right)\right]\right\}^{2} \ $ we can assume $\Re\left(\mathbf{v}^{{\{l\}}}\right)$ with a normal distribution with mean $\mu_{\Re\left(\mathbf{v}^{{\{l\}}}\right)}$ and variance $\sigma_{\Re\left(\mathbf{v}^{{\{l\}}}\right)}^2$. Note that, it is an adequate assumption for a large dataset, due to the central limit theorem. Thus
\begin{multline}
\label{eq:comp_phi_square}
    \mathrm{E}\left\{ \exp{\left[-\Re\left(v_i^{\{l\}}\right)\right]}\right\}= \frac{1}{\sqrt{2\pi\sigma_{\Re\left(\mathbf{v}^{\{l\}}\right)}^2}}\\
    \times\int_0^\infty\exp{\left[-\Re\left(v_i^{\{l\}}\right)\right]}\psi\left[\Re\left(v_i^{\{l\}}\right)\right]\,\mathrm{d}\Re\left(v_i^{\{l\}}\right),
\end{multline}
where $\psi\left(\cdot\right)$ is
\begin{equation}
\psi\left[\Re\left(v_i^{\{l\}}\right)\right]= \exp{\left\{-\frac{\left[\Re\left(v_i^{\{l\}}\right)-\mu_{\Re\left(\mathbf{v}^{\{l\}}\right)}\right]^2}{2\sigma_{\Re\left(\mathbf{v}^{\{l\}}\right)}^2}\right\}}.
\end{equation}

Solving \eqref{eq:comp_phi_square}, yields
\begin{multline}
\label{eq:comp_phi_square_second}
\mathrm{E}\left[ \Re \left( \mathbf{v}^{\{l\}}\right)\right] =\frac{1}{2}\exp\left[\frac{\sigma _{\Re \left(\mathbf{v}^{\{l\}}\right)}^{2}}{2} -2\mu _{\Re \left(\mathbf{v}^{\{l\}}\right)}\right]\\
\times \left[\mathrm{erf}\left(\frac{\mu _{\Re \left(\mathbf{v}^{\{l\}}\right)} -2\sigma _{\Re \left(\mathbf{v}^{\{l\}}\right)}^{2}}{\sqrt{2\sigma _{\Re \left(\mathbf{v}^{\{l\}}\right)}^{2}}}\right) +1\right] ,
\end{multline}
in which $\mathrm{erf}(\cdot)$ is the error function.

Relying on the results of Appendix~\ref{app:2}, we consider that $\mu_{\Re\left(\mathbf{v}^{{\{l\}}}\right)} \gg \sigma_{\Re\left(\mathbf{v}^{\{l\}}\right)}^2$, which is acceptable when the center vectors of $\boldsymbol{\Gamma}^{\{l\}}$ are chosen near to $\mathbf{y}^{\{l-1\}}$. Then, \eqref{eq:comp_phi_square_second} can be simplified to
\begin{equation}
\label{eq:mean_y_sixth}
    \mathrm{E}\left\{ \exp{\left[-\Re\left(\mathbf{v}^{\{l\}}\right)\right]}\right\}=\exp{\left(\frac{\sigma _{\Re \left(\mathbf{v}^{\{l\}}\right)}^{2}}{2} -\mu _{\Re \left(\mathbf{v}^{\{l\}}\right)}\right)},
\end{equation}
similarly
\begin{equation}
\label{eq:var_y_sixth}
    \mathrm{E}\left\{ \exp{\left[-\Re\left(\mathbf{v}^{\{l\}}\right)\right]}^2\right\}=\exp{\left(2\sigma_{\Re\left(\mathbf{v}^{\{l\}}\right)}^2-2\mu_{\Re\left(\mathbf{v}^{\{l\}}\right)}\right)}.
\end{equation}
Then, 
\begin{multline}
    \text{Var}\left\{\exp\left[ -\Re \left( \mathbf{v}^{\{l\}}\right)\right]\right\} =  \exp{\left(2\sigma_{\Re\left(\mathbf{v}^{\{l\}}\right)}^2-2\mu_{\Re\left(\mathbf{v}^{\{l\}}\right)}\right)} \\
    - \exp{\left(\frac{\sigma _{\Re \left(\mathbf{v}^{\{l\}}\right)}^{2}}{2} -\mu _{\Re \left(\mathbf{v}^{\{l\}}\right)}\right)}^2,
\end{multline}
which can be simplified to
\begin{equation}
\label{eq:var_phi}
    \sigma _{{\boldsymbol{\upphi }} ^{\{l\}}}^{2} = {\sigma^2_{\mathbf{v}^{\{l\}}}} \exp\left({-2\mu_{\mathbf{v}^{\{l\}}}}\right).
\end{equation}

Replacing \eqref{eq:var_phi} into \eqref{eq:var_y_IWphi}, results in
\begin{equation}
\label{eq:y_almost_end}
    \sigma _{\mathbf{y}^{\{l\}}}^{2} = I^{\{l\}} \sigma _{\mathbf{w}^{\{l\}}}^{2} {\sigma^2_{\mathbf{v}^{\{l\}}}} \exp\left({-2\mu_{\mathbf{v}^{\{l\}}}}\right).
\end{equation}

Finally, replacing  \eqref{eq:var_v_final} into \eqref{eq:y_almost_end}, results in
\begin{equation}
    \sigma _{\mathbf{y}^{\{l\}}}^{2} =\frac{12}{5} c_{\sigma }^{-2}\exp( -2\mu _{\mathbf{v}^{\{l\}}}) I^{\{l\}}{O^{\{l-1\}}} \sigma _{\mathbf{w}^{\{l\}}}^{2} \sigma _{\upgamma^{\{l\}}}^{4}.
\end{equation}

\bibliographystyle{myIEEEtran.bst}
\bibliography{references.bib}

\end{document}